\documentclass{epsconf}
\usepackage{graphicx}
\usepackage{wrapfig}
\usepackage{amsmath,amssymb}

\title{The electric field of an electron in a electron-hole plasma with degenerate electrons
 }
\author{ \underline{S. P. Sadykova}$^1$, A. A. Rukhadze$^2$}
\institute{$^1$ Forschungszentrum Julich, J{\"u}lich Supercomputing Center, J{\"u}lich, Germany\\
$^2$ Prokhorov General Physics Institute,
Russian Academy of Sciences, Moscow, Russia
}

\begin{document}
\maketitle

\begin{abstract}
We consider the possibility of formation of a superconductivity state either in a semiconductor or in a electron-hole plasma with the degenerate electrons due to the attractive forces between the electrons as a result of the exchange effects through the electron-hole sound wave by analogy to the phonon waves in a solid state. We have determined the view of an interaction potential between two electrons in a degenerate electron-hole plasma. The potential appears to be attractive at distances large than the Debye radius and decreases as $1/r^3$. We discuss the conditions at which the bound electron state - Cooper Pair in a such field can be formed.
\end{abstract}
\section{Introduction}
\label{intro}

\indent Here we would like to consider the conditions for existence of the electron-hole sound waves
in a semiconductor at cryogenic temperatures when electrons (light and of a negative charge) are degenerate and one of the major unsolved problems of the superconductivity theory is determination of the static potential of a point electron. Obviously, the sign of the electron potential in a superconductive medium must be opposite to that in a vacuum. This follows from the Bardeen-Cooper-Schrieffer theory, since the Cooper electron pair can form only when the potential appears to be attractive for both electrons. Below, we will show that in a electron-hole plasma of a semiconductor when the electrons are degenerate and holes are not degenerate, i.e. when

\begin{equation}\label{1}
{\epsilon_{F-}}=\frac{P_{F-}^2}{2m_-}=\frac{{3\pi^2}^{2/3}\hbar^2{n_-}^{2/3}}{2m_-}>>k_BT_-\geq k_BT_+>>\frac{{3\pi^2}^{2/3}\hbar^2{n_+}^{2/3}}{2m_+},
\end{equation}

 a weakly damped electron-hole sound wave with the speed
\begin{equation}\label{2}
{V_{F-}}=\sqrt\frac{2\epsilon_{F-}}{m_-}>>v_s=\sqrt\frac{\epsilon_{F-}}{3m_+}>>{V_{T+}}=\sqrt\frac{k_BT}{m_+}
\end{equation}

can form in such a plasma .
The electron-hole sound wave can be described by the following dispersion relation \cite{1}.
\begin{equation}\label{3}
{\epsilon^{l}(\omega,k)}=1+\frac{3\omega_{L-}^2}{k^2V_{F-}^2}(1+\imath\frac{\pi\omega}{2kV_{F-}})-\frac{\omega_{L+}^2}{\omega^2}=0.
\end{equation}
where the following oscillation spectrum comes from:
\begin{equation}\label{4}
\omega^2=\frac{\omega_{L+}^2}{1+3\omega_{L-}^2/k^2V_{F-}^2}, \:\:\:\:\: \delta=-\frac{3}{4}\frac{\pi m_{+}}{m_-}\frac{\omega_{L+}^4}{k^3 V_{F-}^3}.
\end{equation}
In a long range limit at $3\omega_{L-}^2>>k^2V_{F-}^2$ or 
\begin{equation}\label{5}
k^2r_D^2=k^2 \frac{V_{F-}^2}{3\omega_{L-}^2}<<1.
 \end{equation}
Taking into account that $n_-=n_+$, we can get the linear spectrum from \label{4}.
\begin{equation}\label{6}
\omega=k\sqrt{\frac{n_+}{3n_{-}}\frac{m_-}{m_+}}V_{F-}, \:\:\:\:\: \delta=-\frac{\pi}{4}\frac{n_+}{n_{-}}\sqrt{\frac{n_+}{3n_{-}}\frac{m_-}{m_+}}\omega.
\end{equation}
In a short range limit, when the inverse to \label{5} relation is satisfied, we get
\begin{equation}\label{7}
\omega^2=\omega_{L+}^2, \:\:\:\:\: \delta=-\frac{3}{4}\frac{\pi m_{+}}{m_-}\frac{\omega_{L+}^2}{k^3 V_{F-}^3}.
\end{equation}
\section{The interaction potential of two electrons in a electron-hole plasma}
\label{sec:1}
 The interaction potential of two electrons $\alpha$ and $\beta$ in a electron-hole plasma can be described by the following equation \cite{2}:
\begin{equation}\label{8}
U(r)=\int e^{\imath \vec k\vec r}U(\vec k)d\vec k, \:\:\:\ U(k)=\frac{e_\alpha e_\beta}{2\pi^2}\frac{1}{k^2\varepsilon^l(kV_\alpha,k)}.
\end{equation}

According to the Eq. (ref{3})
\begin{equation}\label{9}
k^2{\epsilon^{l}(kV_\alpha,k)}=k^2+\frac{3\omega_{L-}^2}{V_{F-}^2}-\frac{\omega_{L+}^2}{V_\alpha^2}+\imath \beta, \:\:\: \beta=\frac{3\pi V_\alpha \omega_{L-}^2}{V_{F-}^3},
\end{equation}

here $V_\alpha$ is the speed of a test electron with the charge $e_\alpha$ producing the potential $\phi_\alpha$ at a point $r=0$ where the charge $e_\beta$ is located. Taking into account that $d\vec r /dt=V_\alpha$, it follows $\vec r \uparrow \uparrow V_\alpha$, $V_\alpha>V_{T+}$. As a result, at $e_\alpha=e_\beta=e$ we will get
\begin{equation}\label{10}
  U(k)=\frac{e^2}{2\pi^2}\frac{1}{k^2+\frac{3\omega_{L-}^2}{V_{F-}^2}-\frac{\omega_{L+}^2}{V_\alpha^2x^2}+\imath \beta}, \:\:\:\ U(r)=\frac{e^2}{2\pi^2}\int \frac{e^{\imath  k r x}d\vec k}{k^2+\frac{3\omega_{L-}^2}{V_{F-}^2}-\frac{\omega_{L+}^2}{V_\alpha^2 x^2}+\imath \beta},
\end{equation}
where $x=cos(\theta)$.\\
Let us analyse this expression in two limitting cases: 1) In a short range limit at $r\sim \frac{1}{k}\leq \frac{V_{F-}}{\omega_{L-}}<< \frac{V_\alpha}{\omega_{L+}}\sim \frac{V_{F-}}{\omega_{L+}}$ 2) In a long range limit at $r\sim \frac{1}{k}>> \frac{V_{F-}}{\omega_{L-}}$.
\subsection{The short range limit interaction}



\label{sec:11}
\begin{equation}\label{12}
  U(r)=\frac{e^2}{2\pi^2}\int \frac{e^{\imath \vec k\vec r}d\vec k}{k^2+\frac{1}{r_{D-}^2}}=\frac{e^2}{r}e^{-r/r_{D-}},
\end{equation}
where $r_{D-}=\sqrt{\frac{V_{F-}^2}{3\omega_{L-}^2}}$
\subsection{The long range limit interaction}
\begin{equation}\label{12}
  U(r)=\frac{e^2r_{D-}^2}{2\pi^2}\int \frac{e^{\imath  k r x}d\vec k}{1-\frac{A}{x^2}+\imath\beta_1},
\end{equation}
where $\beta_1=\beta r_{D-}^2>0$, $A=\frac{\omega_{L+}^2}{3\omega_{L-}^2}\frac{V_{F-}^2}{V_\alpha^2}\sim \frac{m_+}{m_-}<<1$.

Taking into account that
$$\lim_{\beta_1\to 0}\frac{1}{1-\frac{A}{x^2}+\imath\beta_1}=-\imath\pi\delta(1-\frac{A}{x^2})$$

 the potential (\ref{12}) will get the following view
\begin{equation}\label{13}
\begin{gathered}
  U_D(r)=-\imath\pi\frac{e^2r_{D-}^2}{\pi}\int_0^\infty k^2 dk\int_{-1}^{+1}dx e^{\imath k r  x}\delta(1-\frac{A}{x^2})=\hfill\\
	-\imath e^2r_{D-}^2\int_0^\infty k^2 dk [ \frac{e^{\imath k r x_0}}{2/x_0}-\frac{e^{-\imath  k  r x_0}}{{2/x_0}}]=\hfill\\
	-\imath\frac{e^2r_{D-}^2 x_0}{2}2\imath\int_0^\infty sin(k r x_0)k^2 dk=\hfill\\
	\frac{e^2r_{D-}^2}{r^3A}\int_0^\infty sin(x)x^2 dx=-2\frac{e^2r_{D-}^2}{r^3A}
		\end{gathered}
\end{equation}

\subsection{Analysis}
It can be clearly seen that at a long range distances the potential has an opposite sign and decays slowly as $-1/r^3$ compared to that.
\begin{figure}[tbp]
\includegraphics[width=0.6\linewidth, height=8cm]{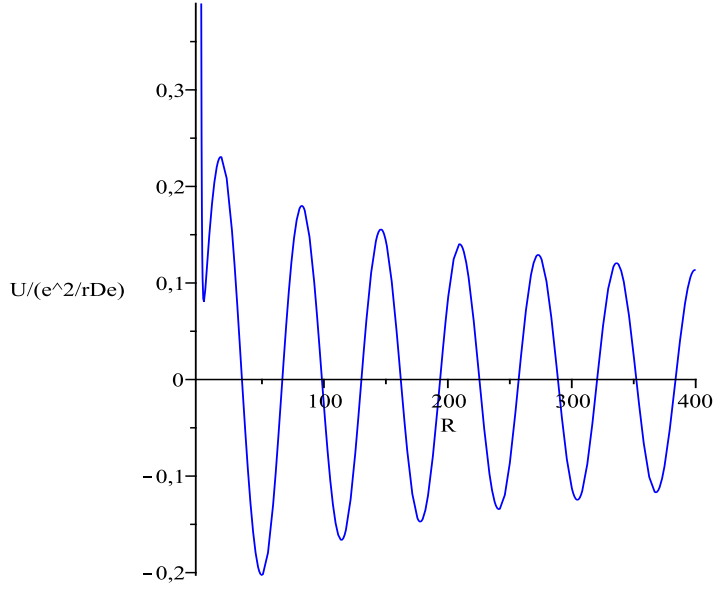}
\caption{The sum of two potentials  (\ref{13}) and the potential (\ref{14}) of two electrons in a  degenerate electron-hole plasma, here $R=r/r_{De}$}
\label{fig:1}
\end{figure}

In a general case the Eq.(\ref{8}) can be deduced to the following integral
\begin{equation}\label{14}
\begin{gathered}
  U(r)=\frac{e^2r_{D-}^2 \sqrt{A}}{r^3}\int_0^\infty \frac{y^2 dy}{({y^2\frac{r_{D-}^2}{r^2}+1})^{3/2}}sin (\frac{y\sqrt A}{\sqrt{y^2\frac{r_{D-}^2}{r^2}+1}})=\frac{e^2r_{D-}^2 \sqrt{A}}{r^3}\int_0^{R\cdot \frac{r_{D-}}{r_{D+} }} \frac{y^2 dy}{({\frac{y^2}{R^2}+1})^{3/2}}sin (\frac{y\sqrt A}{\sqrt{\frac{y^2}{R^2}+1}}) \hfill \\= \frac{e^2r_{D-}^2 }{r^3}\int_0^{\sqrt{A} r/r_D}\frac{z^2 sin(z) dz }{(A-z^2\frac{r_{D-}^2}{r^2})^{5/2}} 
	\end{gathered}
\end{equation}

Having solved the integral (\ref{14}), we get the following from of the potential shown in Fig.\ref{fig:1}, where the integration till the $k\leq 1/r_{Di}$ was performed. Here, we see the multiple minima revealing the quantum nature of the phenomenon: many energy levels, see the Table 1. Here, for the existence of the energy level the following condition must be satisfied :
\begin{table}
\begin{tabular}{l*{8}{c}r}
 MIN & $R_1$& $R_2$ & $ R_{min}$ & $U(R_{min})/U_D$ & $U(R_{min})$, \; eV  & ${U_{min}}_L$, \; eV & ${U_{min}}_L$, \; K & $U(R_{min})$, \; K  \\
\hline
                 1 & 33.85 & 66.36 & 50.46 & -0.202 & -0.053 & $6*10^{-6}$ & 0.07 & 622  \\
2& 98.41 & 130.06 & 114.17 & -0.166 & -0.044 & $6.47*10^{-6}$ & 0.0751 & 511  \\
3& 161.95 & 193.66 & 177.65 & -0.147 & -0.039 & $6.45*10^{-6}$ & 0.0749 & 453  \\
4& 225.35 & 257.12 & 240.88 & -0.134 & -0.035 & $6.4*10^{-6}$ & 0.0746 & 413  \\
5& 288.68 & 320.3 & 304.71 & -0.125 & -0.033 & $6.4*10^{-6}$ & 0.075 & 385  \\
\end{tabular} 
\caption{}
\end{table}
\begin{equation}
U(R_{min})>{U_{min}}_L=\frac{\pi^2\hbar^2}{4m_e a^2},
\end{equation}
here $a$ - potential well width, ${U_{min}}_L$ - its height \cite{L}.

Let us estimate the energy levels of particles moving in a series of potential wells shown in Fig. 
The energy levels in a rectangular potential wells:
\begin{equation}
{{E_{min}}_n}=\frac{\pi^2\hbar^2}{2m_e a^2}n^2, n=1,2 \cdots, N
\end{equation}
where the following condition must be satisfied $U_max>{{E_{min}}_n}$, here 
$S=(R_2-R_1)\cdot U_max$, $S$ - area of the rectangular well with the width $R_2-R_1$ and height $U_{max}$.\\
\indent In the following Table 2 the estimated energy levels are given for a series of potential wells.
\begin{table}
\begin{tabular}{l*{8}{c}r}
 MIN & $R_1$& $R_2$ & $ U_{max}, \; eV$ &$n_1$&  ${E_{min}}_{n_1}, \;eV$ &$n_N$ & ${E_{min}}_{n_N}, \;eV$ \\
\hline
 1 & 33.85 & 66.36 & -0.03235 & 1&$-1.228\cdot 10^{-5}$&51& $-51^2\cdot 1.228\cdot 10^{-5} $ \\
2& 98.41 & 130.06 &- 0.0267 &1&$- 1.2956\cdot 10^{-5}  $&45& $-45^2\cdot 1.2956\cdot 10^{-5}$\\
3& 161.95 & 193.66 & -0.0236&1& $-1.29\cdot 10^{-5} $& 42&$-42^2\cdot 1.29\cdot 10^{-5}$\\
4& 225.35 & 257.12 & -0.0215 & 1&$-1.2859\cdot 10^{-5} $& 40& $-40^2\cdot 1.2859\cdot 10^{-5}$\\
5& 288.68 & 320.3 &1&- 0.01997& $1.298\cdot 10^{-5}$&39&$ -39^2\cdot 1.298\cdot 10^{-5}$   \\
	$\cdot$& $\cdot$& $\cdot$& $\cdot$& $\cdot$ & $\cdot$& $\cdot$ & $\cdot$ \\
45&2812.6  &2844.12  &0.0072&1  &$1.306\cdot 10^{-5} $& 23&$-23^2\cdot 1.306\cdot 10^{-5}$ \\
$\cdot$& $\cdot$& $\cdot$& $\cdot$& $\cdot$ & $\cdot$ & $\cdot$ & $\cdot$\\
120&7538.75  &7570.26  &0.00362&1  &$1.307\cdot 10^{-5}$& 16&$-16^2\cdot 1.307\cdot 10^{-5}$ 
\end{tabular} 
\caption{}
\end{table}

\section{Discussion of the results and quantitative estimations}
\indent From the made above analysis, we can conclude that interaction between two electrons at the distances much larger than the Debye electron radius changes its sign and becomes attractive. This potential amplitude is much higher than that, repulsive one, at the short distances and can lead to the bound state of two electrons (Cooper pair).\\
I would like to express my deepest gratitude and appreciation to Prof. Dr. A. A. Rukhadze for such a long fruitful collaboration with me and personal support.

\end{document}